\documentstyle[12pt]{article}

\def\A{{\cal A}}
\def\B{{\cal B}}
\def\H{{\cal H}}
\def\g{{\gamma}}
\def\l{{\lambda}}
\def\M{{\cal M}}
\def\N{{\cal N}}
\def\P{{\cal P}}
\def\Pp{{\cal P_+}}
\def\Pr{{\cal P_+^\uparrow}}

\def\tPp{{\tilde{\cal P}_+}}
\def\tPr{{\tilde{\cal P}_+^\uparrow}}
\def\RR{{\bf R}}
\def\T{{\cal T}}

\def\bw{{\overline{w}}}
\def\z{{\zeta}}
\def\bz{{\overline{z}}}
\def\sa{{\rm sa}}
\def\Re{{\rm Re}\>}
\def\Im{{\rm Im}\>}

\def\ip<#1|#2>{\left<#1\vphantom{#2}\right|\left.\vphantom{#1}#2\right>}
\def\<#1|#2>{\left<\right.#1\left|\right.#2\left.\right>}
\def\(#1){\left(#1\right)}
\def\[#1|#2]{\left\lbrace #1\vphantom{#2}\right|
             \left.\vphantom{#1}#2\right\rbrace}

\begin{document}
\date{October 1995}
\title{Modular Covariance and the Algebraic PCT/Spin-Statistics Theorem}
\author{D.~R.~Davidson*\\
Dipartimento di Matematica,\quad Universit\`a di Roma ``La Sapienza''\\
00185 Roma, Italy\qquad Email:  davidson@mat.uniroma1.it}
\maketitle
\begin{abstract}
In the theory of nets of observable algebras, the modular operators associated
with wedge regions are expected to have a natural geometric action, a
generalization of the Bisognano-Wichmann condition for nets associated with
Poincar\'e-covariant fields.  Here many possible such modular covariance
conditions are discussed (in spacetime of at least three dimensions),
including several conditions previously proposed and known to imply versions of
the PCT and spin-statistics theorems.  The logical relations between these
conditions are explored:  for example, it is shown that most of them are
equivalent, and that all of them follow from appropriate commutation relations
for the modular automorphisms alone.  These results allow us to reduce the
study of modular covariance to the case of systems describing non-interacting
particles.  Given finitely many Poincar\'e-covariant non-interacting particles
of any given mass, it is shown that modular covariance and wedge duality must
hold, and the modular operators for wedge regions must have the
Bisognano-Wichmann form, so that the usual free fields are the only
possibility.  For models describing interacting particles, it is shown that if
they have a complete scattering interpretation in terms of such non-interacting
particles, then again modular covariance and wedge duality must hold, and the
modular operators for wedge regions must have the Bisognano-Wichmann form, so
that wedge duality and the PCT and spin-statistics theorems must hold.

\bigskip
{\noindent * Supported by a fellowship from the Consiglio Nazionale delle
Ricerche.\\
\noindent Running title:  Modular Covariance}
\end{abstract}
\newpage

\section{Introduction}

Among the most important results of the axiomatic formulation of quantum field
theory are the proofs of the existence of a PCT operator \cite{Jo}, and of the
connection between spin and statistics \cite{LZ,Bg}.  However, these results
rely on complicated analytic continuation arguments that depend heavily on the
detailed structure of Wightman fields.  It would be strange if such highly
physical properties did not have a simpler and more general proof.  Here we
present one such proof, actually a demonstration of the stronger property of
modular covariance, for nets of local algebras satisfying asymptotic
completeness with certain restrictions on their particle content.

It was the work of Bisognano and Wichmann \cite{BW1,BW2} that first introduced
the notion that there should be a geometric interpretation attached to the
modular conjugation and automorphisms with respect to the vacuum for the
algebras of operators associated with certain highly symmetric spacetime
regions.  They worked with algebras associated with a complementary pair of
wedge-shaped
regions, within the context of a set of finite-component
Lorentz-covariant Wightman fields.  In this setting they showed that duality
must hold for such a pair of algebras, that the corresponding modular
automorphisms with respect to the vacuum must be the velocity transformations
that leave the wedge invariant, and finally that the modular conjugations must
be antiunitary reflections---essentially versions of the PCT operator, but with
parity replaced by a reflection appropriate to the wedge.  The property of
duality for such regions, known as wedge duality, implies essential duality for
the corresponding net of local algebras.

There has been a great deal of interest recently in abstracting these notions
to nets of local algebras not necessarily associated with any Wightman field,
for they seem to encode many of the desirable properties of fields in a more
physically direct manner.  In particular, they imply versions of the PCT and
spin-statistics theorems (in the Bisognano-Wichmann theory, by contrast, the
PCT operator is taken as a necessary input).  This has been spurred by
the proof due to Borchers that a weaker result, the covariance of the
translations under modular conjugations and automorphisms, holds under very
general assumptions \cite{Bo1}.  This was then followed by a number of related
results \cite{Wb1,Wb2,Wb3,BS} concerning the interrelations of the modular
structure and the translations, for the most part summarized in \cite{Bo2}.  In
two spacetime dimensions, these are the only relationships required; also for
conformally covariant nets they imply everything desired \cite{FG,BGL1,Wb4,FJ}.
We are concerned here, however, with the remaining cases, for which it is still
not clear what can be proved and what must be assumed.

The Bisognano-Wichmann conditions cannot hold generally as they stand, for
it is easy to construct counterexamples using infinite-component fields
\cite{St,La}.  However, the essence of these examples is that one is free to
specify a representation of the Poincar\'e group for which the
Bisognano-Wichmann conditions do not hold.  The modular operators retain their
geometric interpretation, but they generate a different representation of the
Poincar\'e group.  For this reason one wishes to use a criterion that is
independent of any specified representation of the Poincar\'e group:  one that
simply describes the geometric interpretation of the action of the modular
structure on the local structure of a given net.  Since both structures are
somewhat complicated, there are a number of such criteria that might be and
have been proposed, and an even wider variety of names for them.  The first
goal of this paper is to set out some of these criteria, which we will refer
to generically as relations of modular covariance, and to clarify their
interrelationships.  In particular, we are concerned with the two papers
\cite{GL2} and \cite{Ku1}, which derive related results from somewhat different
modular covariance conditions.  Here (in Theorem 7) we show that under natural
assumptions these conditions (and many others) are all equivalent, and that all
of them in fact follow from much weaker modular covariance premises, ones which
can be expressed entirely in terms of the modular structure, without reference
to the precise local structure of the net.  This we regard as essential to
further study of the possible modular structures of nets.  As we will see, it
allows us to reduce this to the study of nets without interaction.

The modular covariance conditions of Theorem 7 imply the existence of a PCT
operator in even spacetime dimensions.  As stated, however, they apply
only to observable nets, for which they imply that the spin must be integral.
For the full spin-statistics theorem it is necessary to extend the assumptions
and modular covariance conditions slightly to cover field nets containing both
bosonic and fermionic quantities, with normal commutation relations
(Theorem 7$'$).  These results are sufficiently general as to justify our
calling a net `modular covariant' if and only if it satisfies the conditions of
Theorem 7$'$.  This implies the existence of a natural representation of the
Poincar\'e group (and, in even spacetime dimensions, a PCT operator),
determined entirely by the modular structure, under which the field net is
automatically covariant, and for which the Bisognano-Wichmann and
spin-statistics relations hold.  In addition, modular covariance implies wedge
duality, the strongest duality condition that can be expected under these
circumstances.  It is also known that, under rather mild conditions, modular
covariance for the observable net implies modular covariance for the field net,
but we discuss this only briefly here.

We then turn to the study of certain nets without interaction, those described
in terms of a one-particle space by means of Weyl operators.  For these nets
the entire structure is determined by the restriction of the modular operators
to the one-particle space, which determines the localization properties of
the one-particle states.  If modular covariance holds on the one-particle
space, then it holds for the entire net, which then arises from free fields
of the usual sort; on the other hand, if modular covariance does not hold
on the one-particle space, then the net cannot arise from a set of fields.
We show that if the one-particle space carries a physically reasonable
(positive energy, finite spin, finite multiplicity for each mass)
representation of the Poincar\'e group, and if certain standard properties are
obeyed, then modular covariance must hold.  In these cases the usual
one-particle localization is unique, and all such nets arise from free fields.
The representation of the Poincar\'e group is unique, and the
Bisognano-Wichmann condition holds.  The PCT operator, however, is determined
only up to unitary equivalence.

Finally, we consider Poincar\'e-covariant nets having a complete asymptotic
particle interpretation in terms of non-interacting particles of the sort just
described.  Scattering theory in this case is well developed, at least provided
all particles have discrete positive masses, and we now have the additional
information that the asymptotic particles can be described only by free fields.
We are able to adapt some previous results \cite{La} to show that there is a
close relationship between the modular operators for the in-fields, the
out-fields, and the interacting net:  they all satisfy modular covariance with
respect to the unique representation of the Poincar\'e group, differing only in
their choice of PCT operators, and this difference describes the scattering.
Thus we see that the Bisognano-Wichmann condition, wedge duality, the PCT
theorem, and the spin-statistics theorem hold not only for nets associated with
fields, but in addition for all nets with reasonable (massive) scattering
behavior.

\section{Notions of Modular Covariance}

We begin with Minkowski space $\M$ of $d$ spacetime dimensions, with
coordinates $x=(x_0,x_1,\ldots,x_{d-1})$, where $x_0$ is the time coordinate.
We will in general assume $d>2$, since although some of our results hold also
in the lower-dimensional cases, they are no longer particularly relevant there.
For the sake of concreteness, one might simply take $d=3$, since the results
are generally such that if they hold in three dimensions they hold also in
higher dimensions, but we will avoid explicit references to $d$.  The
$d$-dimensional Poincar\'e group $\P$ is the group of all inhomogeneous linear
transformations
of $\M$ preserving the metric with diagonal elements
$(1,-1,\ldots,-1)$.  We will be particularly interested in the subgroups $\Pp$,
the proper Poincar\'e group, generated by the translations and all homogeneous
transformations of determinant $+1$, and $\Pr$, the restricted Poincar\'e
group, consisting of all proper orthochronous Poincar\'e transformations
(those that preserve the sign of the time component).  We will also make use of
their universal covering groups $\tPp$ and $\tPr$.

Within Minkowski space we distinguish the family of wedge regions:  the
particular complementary pair of wedges $W_R=\[x|x_1>{|x_0|}]$ and
$W_L=\[x|x_1<-{|x_0|}]$ and their Poincar\'e transforms.  (We will think of
$W_L$ as complementary to $W_R$, and write $W_R'=W_L$, even though strictly
speaking $W_R'=\overline{W_L}$).  A complementary pair of wedges has a common
vertex (for $W_R$ and $W_L$, the hyperplane $x_1=x_0=0$) and opposite sets of
directions (for $W_R$, a set of directions including $\hat x_1$; for $W_L$, a
set including $-\hat x_1$).  We will write $W_1\|W_2$ if the vertices of $W_1$
and $W_2$ are parallel, specialized as $W_1\|_s W_2$ if in addition they have
the same directions, or $W_1\|_a W_2$ if they have opposite directions.

To each wedge $W$ we associate certain Poincar\'e transformations (which we
will actually use primarily as maps of the family of wedges):  a reflection
$j(W)$ about the vertex of the wedge, and a one-parameter family of velocity
transformations (in the appropriate reference frame, and with an appropriate
scale) $\l(W,t)$ in the direction of the wedge, both leaving the vertex fixed.
For example,
$j(W_R)(x_0,x_1,x_2,\ldots,x_d)=(-x_0,-x_1,x_2,\ldots,x_d)$, and
$\l(W_R,t)(x_0,x_1,x_2,\ldots,x_d)=(x_0',x_1',x_2,\ldots,x_d)$, with
$x_0'=x_0\cosh 2\pi t+x_1\sinh 2\pi t$ and
$x_1'=x_1\cosh 2\pi t+x_0\sinh 2\pi t$.  In general, $j(W)$ and $\l(W,t)$ are
conjugates of these within the Poincar\'e group.  Thus $j(W)$ is a proper but
time-reversing involutory Poincar\'e transformation, which interchanges $W$
with $W'$, while the $\l(W,t)$ form a one-parameter group of proper
orthochronous Poincar\'e transformations, leaving $W$ and $W'$ invariant, and
such that $\l(W',t)=j(W)\l(W,t)j(W)=\l(W,-t)$.  The $\l(W,t)$ generate the
entire restricted Poincar\'e group; since $d>2$, the $j(W)$ generate the entire
proper Poincar\'e group.

For our purposes, the only data required from a net will be a map from the
family of wedges $W$ to a family $\A(W)$ of von Neumann algebras of operators
on a Hilbert space $\H$, with a distinguished vacuum vector $\Omega$.  The
conclusions drawn will also apply directly only to the wedge algebras; if these
results are to be applied to a local net, then to begin with it must satisfy
essential duality, and the statements must be considered as referring to its
dual net.  In what follows we will make the following assumptions:
\goodbreak\noindent
(i) $\A(W')\subset\A(W)'$ (locality);\\
(ii) $\A(W_1)\subset\A(W_2)$ whenever $W_1\subset W_2$ (isotony);\\
(iii) $\Omega$ is a cyclic vector for $\A(W_1)\cap\A(W_2)$ whenever
$W_1\cap W_2\neq\emptyset$ (cyclicity);\\
(iv) for all $W_1$ and $W_2$, $\A(W_1)=\[A(W_1)\cap\A(W)|{W\|_s W_2}]''$
(an additivity property).

If the wedge algebras are in fact derived from a local net, then (i)--(iv)
follow from standard assumptions, but we prefer to state them here in the form
required.  We will show in many cases that a stronger version of (i) actually
holds, namely $\A(W')=\A(W)'$ (wedge duality).  This implies essential duality,
so that if the wedge algebras are derived from a local net, then there is some
maximal local net consistent with them, which satisfies duality.

Assumptions (i)--(iv) imply, among other things, that for every wedge $W$,
there are modular involution, conjugation, and automorphism operators $S(W)$,
$J(W)$ and $\Delta(W)$ for the pair of algebras $\A(W),\A(W)'$ with respect
to the vacuum $\Omega$.  They may be defined by the unique polar decomposition
$S(W)=J(W)\Delta(W)^{1/2}$ of the closed antilinear operator $S(W)$, where
$S(W)$ is defined such that $S(W)X\Omega=X^*\Omega$ for every $X\in\A(W)$.
Then $S(W)\psi=\psi$ if and only if $\psi\in\overline{\A(W)^\sa\Omega}$.
$S(W)$ is an antilinear involution, $J(W)$ is an antiunitary involution, and
$\Delta(W)$ is self-adjoint and positive.  The modular conjugation is the
adjoint action of $J(W)$, and the modular automorphism group is the
one-parameter group given by the adjoint action of $\Delta(W)^{it}$.  The
properties of these operators and their actions are well known from the
Tomita-Takesaki theory:  for example, $J(W)\Omega=\Delta(W)\Omega=0$ and
$J(W)\Delta(W)J(W)=\Delta(W)^{-1}$; also $J(W)\A(W)J(W)=\A(W)'$, and
$\Delta(W)^{it}\A(W)\Delta(W)^{-it}=\A(W)$ for all real $t$.

Where necessary, we will also make the following additional assumption:\\
(v) $J(\Lambda W)$ is a weakly continuous function of $\Lambda\in\Pr$
(a continuity property).\\
This property too typically holds for most nets that are usually considered;
for example, it follows from the covariance of the net under any strongly
continuous unitary representation of $\Pr$.  For some of our results, however,
it must be specifically assumed.

Then relations of modular covariance will connect the action of the modular
conjugation operators $J(W)$ with the transformations $j(W)$, and the
action of the modular automorphism operators $\Delta(W)^{it}$ with the
transformations $\l(W,t)$.  Let us list a number of possible conditions:

\smallskip\goodbreak\noindent
(a) covariance under modular conjugations of modular conjugations,
\begin{equation}
J(W_1)J(W_2)J(W_1)=J(j(W_1)W_2);
\end{equation}
(b) covariance under modular conjugations of modular automorphisms,
\begin{equation}
J(W_1)\Delta(W_2)J(W_1)=\Delta(j(W_1)W_2);
\end{equation}
(c) covariance under modular conjugations of modular involutions,
\begin{equation}
J(W_1)S(W_2)J(W_1)=S(j(W_1)W_2);
\end{equation}
(d) covariance under modular conjugations of wedge algbras,
\begin{equation}
J(W_1)\A(W_2)J(W_1)=\A(j(W_1)W_2);
\end{equation}
(e) covariance under modular automorphisms of modular conjugations,
\begin{equation}
\Delta(W_1)^{it}J(W_2)\Delta(W_1)^{-it}=J(\l(W_1,t)W_2);
\end{equation}
(f) covariance under modular automorphisms of modular automorphisms,
\begin{equation}
\Delta(W_1)^{it}\Delta(W_2)\Delta(W_1)^{-it}=\Delta(\l(W_1,t)W_2);
\end{equation}
(g) covariance under modular automorphisms of modular involutions,
\begin{equation}
\Delta(W_1)^{it}S(W_2)\Delta(W_1)^{-it}=S(\l(W_1,t)W_2);
\end{equation}
(h) covariance under modular automorphisms of wedge algebras,
\begin{equation}
\Delta(W_1)^{it}\A(W_2)\Delta(W_1)^{-it}=\A(\l(W_1,t)W_2);
\end{equation}
(i) the modular conjugations $J(W)$ are the representatives of $j(W)$ under
some representation of the proper Poincar\'e group;\\
(j) the modular automorphisms $\Delta(W)^{it}$ are the representatives of
$\l(W,t)$ under some representation of the restricted Poincar\'e group.
\bigskip\goodbreak

These ten statements, and their combinations, cover most of possibilities
for modular covariance relations that apply to all combinations of wedges.
(We exclude here those referring only to wedges related in a certain way---for
example, the conditions of modular inclusion in \cite{Bo2}, which apply only to
parallel wedges.)  In Theorem 7, we will demonstrate the equivalence of the
majority of these modular covariance relations, including those of \cite{GL2}
(covariance under modular automorphisms of wedge algebras) and of \cite{Ku1}
(covariance under modular conjugations of wedge algebras).

Note that when we speak of a representation of the Poincar\'e group, we must
require the conditions appropriate to a group of symmetries:  that is, we refer
to a strongly continuous projective representation by unitary or antiunitary
operators.  This will therefore be a representation of the covering group,
either $\tPp$ or $\tPr$, in which the connected component of the identity
$\tPr$ must be represented by unitary operators.  If the representation
condition (i) above is to hold, then the time-reversing operators in $\tPp$
must be represented by antiunitary operators (which is in fact what we would
expect physically, due to the positivity of the energy).  Notice also, for
example, that since $J(W)$ is antiunitary, (b) above implies that
$J(W_1)\Delta(W_2)^{it}J(W_1)=\Delta(j(W_1)W_2)^{-it}$.

\section{Equivalence of Strong and Weak Formulations}

We begin by discussing the relationship between conditions dealing with
the wedge algebras, and those dealing purely with the modular structure.
The modular operators associated with a given algebra contain much less
information than the algebra itself, but there is the following weak result:
if $\psi\in\H$ is such that $S(W)\psi=\psi$, then there is a closed symmetric
operator $\tilde X$ affiliated with $\A(W)$ such that $\tilde X\Omega=\psi$.
It is defined on the core $\A(W)'\Omega$ by $\tilde X Y\Omega=Y\psi$ for all
$Y\in\A(W)'$.  (The following results could of course be proved without this
machinery, but only at the cost of a certain increase in notational complexity;
furthermore, the method we use seems in accord with the modular spirit of
our presentation.)

\bigskip\goodbreak
\noindent{\bf Lemma 1:}~~
{\em
Suppose assumptions (i)--(iv) hold.  Let $U$ be a unitary (or antiunitary)
operator such that $U\Omega=\Omega$, and $\g$ be a Poincar\'e transformation
such that $US(W)U^*=S(\g W)$ for every wedge $W$.  If there is some particular
wedge $W_0$ such that $U\A(W_0)U^*=\A(\g W_0)$, then likewise
$U\A(W)U^*=\A(\g W)$ for every wedge $W$.
}

\smallskip\goodbreak
\noindent{\it Proof:}~~
{
Let us first show that the statement holds when $W\subset W_0$.  Then we have
immediately $U\A(W)U^*\subset U\A(W_0)U^*=\A(\g W_0)$, and also
$\A(\g W)\subset\A(\g W_0)$.  If $X\in\A(W)^\sa$, then
$UXU^*\in\A(\g W_0)^\sa$, and furthermore
$UX\Omega\in\overline{\A(\g W)^\sa\Omega}$.  Thus there is a closed symmetric
operator $\tilde X$ affiliated with $\A(\g W)$ (and thus also with
$\A(\g W_0)$) such that $\tilde X\Omega=UX\Omega$.  But $\tilde X$ and
$UXU^*$ agree on the dense set $\A(\g W_0)'\Omega$, from which it follows that
$\tilde X$ is in fact bounded and equal to $UXU^*$, which therefore lies in
$\A(\g W)$.  Thus $U\A(W)U^*\subset\A(\g W)$.  On the other hand, we may
apply the same argument to $\g W_0$ and $\g W$, with $U$ and $U^*$
interchanged and $\g^{-1}$ taking the role of $\g$, to show that
$U^*\A(\g W)U\subset\A(W)$.  Thus $U\A(W)U^*=\A(\g W)$.

Likewise we have $U\A(W_0)'U^*=\A(\g W_0)'$, so that, by the same reasoning,
if $\A(W)'\subset\A(W_0)'$, we have $U\A(W)'U^*=\A(\g W)'$ and again
$U\A(W)U^*=\A(\g W)$.  Thus the statement of the lemma also holds whenever
$W'\subset W_0'$, i.e. whenever $W\supset W_0$.  If we have merely that
$W\|_s W_0$, then there is some $W_1\subset W_0\cap W$.  The statement of the
lemma holds for $W_1$, and hence we may repeat the argument with $W_1$ in place
of $W_0$ to show that it holds for $W$.  Thus $U\A(W)U^*=\A(\g W)$ whenever
$W\|_s W_0$.

Next let us consider the case in which neither $W\cap W_1$ nor $W'\cap W_1'$ is
empty for any
$W_1\|_s W_0$ (as is the case for most choices of $W$ and $W_0$).
Then the vacuum is cyclic and separating for both $\A=\A(W)\cap\A(W_1)$ and
$\B=\A(\g W')\cap\A(\g W_1')$.  By the results above,
$U\A U^*\subset\A(\g W_1)$.  If $X\in\A^\sa$, then $UXU^*\in\A(\g W_1)^\sa$,
but also $UX\Omega\in\overline{\A(\g W)^\sa\Omega}$.  Thus there is a closed
symmetric operator $\tilde X$ affiliated with $\A(\g W)$ such that
$\tilde X\Omega=UX\Omega$.  But $\tilde X$ and $UXU^*$ agree on the dense set
$\B\Omega$, from which it follows that $\tilde X$ is in fact bounded and equal
to $UXU^*$, which therefore is in $\A(\g W)$.  Thus $U\A U^*\subset\A(\g W)$.
Letting $W_1$ vary we generate all of $\A(W)$, so that
$U\A(W)U^*\subset\A(\g W)$.  But again as above we can use a similar argument
to show that $U^*\A(\g W)U\subset\A(W)$, so that in fact $U\A(W)U^*=\A(\g W)$.
For the remaining wedges, we may repeat these arguments to show that the result
holds generally.
}
\bigskip\goodbreak

This becomes useful when combined with the Tomita-Takesaki theorem, as follows:

\smallskip\goodbreak
\noindent{\bf Theorem 2:}~~
{\em
Under assumptions (i)--(iv), the following are equivalent:\\
(a) covariance under modular conjugations of both modular conjugations\\
\hphantom{(a)} and modular automorphisms;\\
(b) covariance under modular conjugations of modular involutions;\\
(c) covariance under modular conjugations of wedge algbras.\\
If these conditions hold, then so also does wedge duality.
}

\smallskip\goodbreak
\noindent{\bf Theorem 3:}~~
{\em
Under assumptions (i)--(iv), the following are equivalent:\\
(a) covariance under modular automorphisms of both modular conjugations\\
\hphantom{(a)} and modular automorphisms;\\
(b) covariance under modular automorphisms of modular involutions;\\
(c) covariance under modular automorphisms of wedge algbras.\\
If these conditions hold, then so also does wedge duality.
}

\smallskip\goodbreak
\noindent{\it Proof:}~~
{
In each case it is clear from the definition of $S(W)$ that (c) implies (b),
and from the uniqueness of the polar decomposition that (b) is equivalent
to (a).  Any of the assumptions of Theorem 2 implies that $J(W')=J(j(W)W)=J(W)$
and $\Delta(W')=\Delta(j(W)W)=\Delta(W)^{-1}$.  Since we already have that
$\A(W')\subset\A(W)'$, this implies that $\A(W')=\A(W)'$.  Likewise in Theorem
3, $\A(W')$ is a subalgebra of $\A(W)'$, whose modular operators are invariant
under the modular automorphism group for $\A(W)'$, and for which the vacuum is
cyclic.  It follows that $\A(W')$ is invariant under the modular automorphism
group for $\A(W)$, and by a standard result it is therefore equal to $\A(W)'$.
It remains for us to show that under assumptions (i)--(iv), (b) implies (c).
This follows from Lemma 1 using the Tomita-Takesaki results
$J(W)\A(W)J(W)=\A(W')=\A(j(W)W)$ and
$\Delta(W)^{it}\A(W)\Delta(W)^{-it}=\A(W)=\A(\l(W,t)W)$.
}
\bigskip

Thus we see that the covariance of wedge algebras can always be reduced to
appropriate statements referring purely to the relations of modular operators
to one another, without reference to the algebras themselves.

\section{Representations of the Poincar\'e Group}

We have now to discuss the relationships between the modular covariance
conditions referring only to the modular structure and those calling for the
existence of certain representations of the Poincar\'e group.  In this we make
use of the results of \cite{BGL2}, which establish the existence of such
representations under rather weak conditions.

\bigskip
\noindent{\bf Theorem 4:}~~
{\em
Under assumptions (i)--(v), the following are equivalent:\\
(a) covariance under modular conjugations of modular conjugations;\\
(b) the modular conjugations $J(W)$ are representatives of $j(W)$ under a
representation of the covering group $\tPp$ of the proper Poincar\'e group;\\
(c) the modular conjugations $J(W)$ are the representatives of $j(W)$ under a
representation of the proper Poincar\'e group $\Pp$ which represents
orthochronous transformations by unitary operators and time-reversing
transformations by antiunitary operators; the vacuum is invariant under it,
and the modular conjugations are covariant, and wedge duality holds.\\
Assumption (v) is not necessary if (b) or (c) holds.
}

\smallskip
\noindent{\it Proof:}~~
{
Clearly (c) implies (b), and (b) implies both (a) and assumption (v).  We
therefore assume (a) and assumptions (i)--(v) and seek to prove the rest.  We
begin with a single one-parameter subgroup $\Lambda(t)$ of the Poincar\'e group
(necessarily the restricted group), for which we suppose that there is some
wedge $W$ such that $j(W)\Lambda(t)j(W)=\Lambda(-t)$.  From this it follows
that $j(\Lambda(t)W)\Lambda(t')W=\Lambda(2t-t')W'$.  (There are many such
instances:  for example, $W_R$ will serve for the translations in the
$\hat x_1$ direction, the velocity transformations in the $\hat x_2$ direction,
or the rotations about $\hat x_3$.  Thus there is such a $W$ if $\Lambda(t)$ is
conjugate to any of these one-parameter subgroups, and in particular such
subgroups generate all of $\Pr$.)

We can then apply the methods of \cite{Bo2}, Proposition 3.1 and Lemma 3.2,
the proof of which can be simplified as follows.  If we write
$J_t=J(\Lambda(t)W)$, then from our assumptions $J_tJ_{t'}J_t=J_{2t-t'}$.
First we wish to show by induction that $J_{nt}J_{(n+1)t}=J_0J_t$ for all
integers $n$.  This holds for $n=0$, but also
$J_{nt}J_{(n+1)t}J_{nt}=J_{(n-1)t}$, so that
$J_{nt}J_{(n+1)t}=J_{(n-1)t}J_{nt}$, and the induction proceeds in either
direction.  Next we wish to show by induction that $(J_0J_t)^n=J_0J_{nt}$ for
all integers $n$.  This is immediate for $n=0,\pm 1$, and
\begin{equation}
(J_0J_t)^{n+1}=(J_0J_t)^nJ_0J_t=J_0J_{nt}J_{nt}J_{(n+1)t}=J_0J_{(n+1)t}
\end{equation}
provides the induction for $n>0$.  But then by the same result
$(J_0J_t)^{-n}=(J_0J_{-t})^n=J_0J_{-nt}$.  From this we have
$J_0J_{nt}J_0J_{mt}=(J_0J_t)^{m+n}=J_0J_{(n+m)t}$, and in general if $t$
and $t'$ are rationally related then $J_0J_tJ_0J_{t'}=J_0J_{t+t'}$.  Then by
continuity it follows that $J_0J_t$ is a continuous one-parameter unitary
group.

Thus also $U(\Lambda(t))=J(W)J(\Lambda(-t/2)W)$ is a continuous one-parameter
unitary group, which can be seen to implement $\Lambda(t)$ on the modular
conjugations:  for an arbitrary wedge $W_1$,
$U(\Lambda(t))J(W_1)U(\Lambda(-t))=J(\Lambda(t)W_1)$.  Likewise $U(\Lambda(t))$
is covariant under the modular conjugations:
$J(W_1)U(\Lambda(t))J(W_1)=U(j(W_1)\Lambda(t)j(W_1))$.  Thus also
the $U(\Lambda(t))$ are covariant with respect to each other:
$U(\Lambda'(t'))U(\Lambda(t))U(\Lambda'(-t'))
=U(\Lambda'(t')\Lambda(t)\Lambda'(-t'))$.  We may then apply the methods of
\cite{BGL2}, as also in \cite{GL2}, Proposition 2.4:  the $U(\Lambda(t))$
generate a central weak Lie extension of $\Pr/H$, where $H$ is the normal
subgroup $H=\[\Lambda\in\Pr|J(\Lambda W)=J(W)\hbox{\rm{ for all }}W]$ of $\Pr$.
If $H$ is trivial or the translation subgroup, then $\Pr/H$ is the restricted
Poincar\'e or Lorentz group, and we have a representation of the corresponding
covering group, hence in either case a representation of $\tPr$.  If $H=\Pr$,
then $J(W_1)=J(W_2)$ for all $W_1,W_2$, and the representation is trivial
immediately.  Thus in any case we have a representation of the covering group
$\tPr$, under which the modular conjugations are covariant.  It is then
straightforward to extend this to a representation of the covering group $\tPp$
of the proper Poincar\'e group, since as we have seen the representation is
also covariant under the modular conjugations.  Since the vacuum is invariant
under each $J(W)$, it is also invariant under this representation.  As we
have remarked before, the subgroup $\tPr$ must be represented by unitary
operators, and the time-reversing transformations by antiunitary operators.

To show that this representation is in fact of the proper Poincar\'e group
itself, we follow the procedure used in \cite{GL2} and \cite{Ku1}:  let
$R(\theta)$ be the representative of the rotation by the angle $\theta$ about
the $\hat x_3$ axis, so that $J(W_R)R(\theta)J(W_R)=R(-\theta)$.  We have
$J(W_R)=J(j(W_L)W_L)=J(W_L)$, so
$1=J(W_L)J(W_R)=R(\pi)J(W_R)R(-\pi)J(W_R)=R(2\pi)$.
}
\bigskip\goodbreak

Note that an extension of a representation of the restricted Poincar\'e
group to one of the proper Poincar\'e group, in which the time-reversing
transformations are represented by antiunitary operators, is almost unique, but
not quite.  One has always a choice of phase---that is, the time-reversing
transformations may always be multiplied by any common unitary operator $V$
which commutes with the restricted Poincar\'e group and anticommutes
($VU=UV^*$) with the time-reversing transformations.  For example, if the
representation is irreducible, then $V$ can be any complex phase
$e^{i\theta}$.

Next we introduce a lemma that allows us to connect the behavior of modular
automorphisms with that of modular conjugations:

\bigskip\goodbreak
\noindent{\bf Lemma 5:}~~
{\em
Suppose assumptions (i)--(iii) hold.  If $W_1, W_2$ are two wedges such that
$W_1\cap W_2\neq\emptyset$ and $W_1'\cap W_2'\neq\emptyset$, then
\begin{equation}
\Delta(W_1)^{1/2}\Delta(W_2)^{-1/2}\subset J(W_1)J(W_2).
\end{equation}
That is, the operator on the left is densely defined and closable, and the
bounded operator on the right extends it.  This implies among other things
that for every $\psi\in D(\Delta(W_1)^{1/2})$, $\phi\in D(\Delta(W_2)^{-1/2})$,
we have
\begin{equation}
\ip<\Delta(W_1)^{1/2}\psi|\Delta(W_2)^{-1/2}\phi>
=\ip<\vphantom{\Delta(W_1)^{1/2}}J(W_1)\psi|J(W_2)\phi>.
\end{equation}

Let $U$ be a unitary (or antiunitary) operator, and $\g$ a Poincar\'e
transformation such that $U\Delta(W)U^*=\Delta(\g W)$ for every wedge $W$.
Then also
\begin{equation}
UJ(W_1)J(W_2)U^*=J(\g W_1)J(\g W_2)
\end{equation}
for any pair of wedges $W_1,W_2$, and there is a unitary operator $V$ such that
$UJ(W)U^*=VJ(\g W)$ for every wedge $W$.  If in addition there is some
particular wedge $W_0$ such that $UJ(W_0)U^*=J(\g W_0)$, then likewise
$UJ(W)U^*=J(\g W)$ for every wedge $W$.
}

\smallskip\goodbreak
\noindent{\it Proof:}~~
{
By assumption, the vacuum is cyclic for $\A=\A(W_1)\cap\A(W_2)$.  If $X\in\A$,
then $S(W_1)X\Omega=S(W_2)X\Omega=X^*\Omega$.  Thus $S(W_1)S(W_2)$ agrees with
the identity on the dense set $\A\Omega$.  But $S(W_1)=J(W_1)\Delta(W_1)^{1/2}$
and $S(W_2)=\Delta(W_2)^{-1/2}J(W_2)$.  Thus the bounded operator
$J(W_1)J(W_2)$ agrees with the product $\Delta(W_1)^{1/2}\Delta(W_2)^{-1/2}$ on
the dense set $J(W_2)\A\Omega$.  Applying the same reasoning to
$\A(W_2)'\supset\A(W_2')$ and $\A(W_1)'\supset\A(W_1')$ shows that
$\Delta(W_2)^{-1/2}\Delta(W_1)^{1/2}$ agrees with $J(W_2)J(W_1)$ on a dense
set.  In particular, $\Delta(W_1)^{1/2}\Delta(W_2)^{-1/2}$ has a densely
defined adjoint, and thus is closable.  Then since
$\Delta(W_1)^{1/2}\Delta(W_2)^{-1/2}$ agrees with $J(W_1)J(W_2)$ on a dense
set, they must agree wherever defined.  Furthermore since
$\<\Delta(W_1)^{1/2}\psi|\Delta(W_2)^{-1/2}\phi>=\<J(W_1)\psi|J(W_2)\phi>$
for a dense set of $\psi$ and $\phi$, and since the right-hand side is a
bounded function of $\psi$ and $\phi$, equality must hold whenever the
left-hand side is defined.

For the second part, let us first assume that $W_1$ and $W_2$ satisfy the
condition of the first
part.  Then $UJ(W_1)J(W_2)U^*$ extends
$U\Delta(W_1)^{1/2}\Delta(W_2)^{-1/2}U^*$, but by assumption the
latter is equal to $\Delta(\g W_1)^{1/2}\Delta(\g W_2)^{-1/2}$, which extends
to the bounded operator $J(\g W_1)J(\g W_2)$.  Thus $UJ(W_1)J(W_2)U^*$ and
$J(\g W_1)J(\g W_2)$ are extensions of the same densely defined closable
operator, and must in fact be equal.  This is so provided that
$W_1\cap W_2\neq\emptyset$ and $W_1'\cap W_2'\neq\emptyset$, but in any case
we can find some $W_3$ such that none of $W_1\cap W_3$, $W_2\cap W_3$,
$W_1'\cap W_3'$, or $W_2'\cap W_3'$ is empty.  Then again we have
\begin{eqnarray}
UJ(W_1)J(W_2)U^*&=&UJ(W_1)J(W_3)U^*UJ(W_3)J(W_2)U^*\nonumber\\
&=&J(\g W_1)J(\g W_3)J(\g W_3)J(\g W_2)=J(\g W_1)J(\g W_2),
\end{eqnarray}
without restriction on $W_1,W_2$.  We may rearrange this to obtain
\begin{equation}
UJ(W_1)U^*J(\g W_1)=UJ(W_2)U^*J(\g W_2)=V
\end{equation}
where $V$ is a single unitary operator independent of the choice of wedges.
Thus $UJ(W)U^*=VJ(\g W)$ for any wedge $W$.  Then if $UJ(W_0)U^*=J(\g W_0)$
we have $VJ(\g W_0)=J(\g W_0)$ and $V=I$.
}
\bigskip\goodbreak

This result is a very strong condition on the modular operators, with
several important consequences.  First, the modular automorphisms must be such
that the product of $\Delta(W_1)^{1/2}$ and $\Delta(W_2)^{-1/2}$ is densely
defined and is the restriction of a unitary operator for every $W_1,W_2$ with
$W_1\cap W_2\neq\emptyset$ and $W_1'\cap W_2'\neq\emptyset$.  Second, the
modular automorphisms almost determine the modular conjugations:  they
determine the products of pairs of modular conjugations, or the modular
conjugations themselves up to a unitary phase operator $V$.  This will be used
several times in the next section, in the proof of our main theorem on modular
covariance.

All that remains is to show that a corresponding condition holds for
representations of the Poincar\'e group, as is necessary if the modular
operators are to be derived from such a representation.

\bigskip\goodbreak
\noindent{\bf Lemma 6:}~~
{\em
Let $U(\lambda)$ be a representation of the restricted Poincar\'e group $\Pr$
satisfying the spectrum condition.  If $W_1$ and $W_2$ are two wedges such that
$W_1\cap W_2\neq\emptyset$ and $W_1'\cap W_2'\neq\emptyset$, and we write
$\Delta_0(W)^{it}=U(\l(W,t))$, then
\begin{equation}
\ip<\Delta_0(W_1)^{1/2}\psi|\Delta_0(W_2)^{-1/2}\phi>
=\ip<\vphantom{\Delta_0(W_1)^{1/2}}\psi|U(j(W_1)j(W_2))\phi>
\end{equation}
for every $\psi\in D(\Delta_0(W_1)^{1/2})$, $\phi\in D(\Delta_0(W_2)^{-1/2})$.
}

\smallskip\goodbreak
\noindent{\it Proof:}~~
{
{}From the proof of Theorem 1.1 of \cite{GL2} we can see that
\begin{eqnarray}
&&\ip<\Delta_0(W_1)^{1/2}\psi|\Delta_0(\l(W,t)W_1)^{-1/2}\phi>\nonumber\\
&&\qquad\qquad=\ip<\Delta_0(W_1)^{1/2}\psi|
\Delta_0(W)^{it}\Delta_0(W_1)^{-1/2}\Delta_0(W)^{-it}\phi>\\
&&\qquad\qquad
=\ip<\vphantom{\Delta_0(W_1)^{1/2}}\psi|U(j(W_1)\l(W,t)j(W_1)\l(W,-t))\phi>
=\ip<\vphantom{\Delta_0(W_1)^{1/2}}\psi|U(\l(W,-2t))\phi>\nonumber
\end{eqnarray}
whenever $W_1$ and $W$ are orthogonal wedges, so that our condition holds
for $W_1$ and $W_2=\l(W,t)W_1$.  The same proof also can be adapted to give
the same result if $W_2=\Lambda(t)W_1$ where $\Lambda(t)$ is any
one-parameter subgroup of the Lorentz group such that $W_1$ and $W_2$
satisfy the conditions of the present lemma.  Thus the lemma holds
whenever $W_1$ and $W_2$ are Lorentz transforms of each other.

If $W_1$ and $W_2$ are arbitrary wedges satisfying the conditions of
our lemma, then there is some $W_3\|_s W_2$ such that $W_1$ and $W_3$ also
satisfy the conditions of our lemma, and in addition are Lorentz transforms of
each other.  Thus there is some translation $T(x)$ having no component parallel
to the vertex of $W_2$, with $W_3$ the translate by $x$ of $W_2$, and
\begin{eqnarray}
&&\ip<\Delta_0(W_1)^{1/2}\psi|\Delta_0(W_2)^{-1/2}\phi>=
\ip<\Delta_0(W_1)^{1/2}\psi|\Delta_0(W_3)^{-1/2}\Delta_0(W_3)^{1/2}
\Delta_0(W_2)^{-1/2}\phi>\nonumber\\
&&\qquad\qquad=
\ip<\Delta_0(W_1)^{1/2}\psi|\Delta_0(W_3)^{-1/2}T(x)\Delta_0(W_2)^{1/2}
T(-x)\Delta_0(W_2)^{-1/2}\phi>\\
&&\qquad\qquad=\ip<\vphantom{\Delta_0(W_1)^{1/2}}\psi|
U(j(W_1)j(W_3))T(x)\Delta_0(W_2)^{1/2}T(-x)\Delta_0(W_2)^{-1/2}\phi>\nonumber
\end{eqnarray}
for all $\psi\in D(\Delta_0(W_1)^{1/2})$ and all $\phi$ for which the
expression is defined.

At this point we employ a converse to Borchers' Theorem (\cite{Da}, Theorem 3),
a consequence of the analytic continuation made possible by the spectrum
condition.  The theorem shows, for example, that for the particular wedge
$W_R$, the expression $T(x)\Delta_0(W_R)^{1/2}T(-x)\Delta_0(W_R)^{-1/2}$ agrees
with $T(2x)$ wherever it is defined, provided $x$ lies in the
$\hat x_0+\hat x_1$ or $\hat x_0-\hat x_1$ directions.  Thus this is true also
for linear combinations of these directions, i.e., in general, whenever $x$ has
no component parallel to the vertex of the wedge.  This means that
\begin{eqnarray}
&&\ip<\vphantom{\Delta_0(W_1)^{1/2}}\psi|
U(j(W_1)j(W_3))T(x)\Delta_0(W_2)^{1/2}T(-x)\Delta_0(W_2)^{-1/2}\phi>\\
&&\qquad\qquad=
\ip<\vphantom{\Delta_0(W_1)^{1/2}}\psi|U(j(W_1)j(W_3))T(2x)\phi>
=\ip<\vphantom{\Delta_0(W_1)^{1/2}}\psi|U(j(W_1)j(W_2))\phi>\nonumber
\end{eqnarray}
for all $\psi,\phi$ for which the expression is defined.  Thus the lemma holds
generally.
}

\section{Modular Covariance}

We are now ready for our main (and rather heavily overloaded) theorem on
modular covariance.

\bigskip
\noindent{\bf Theorem 7:}~~
{\em
Under assumptions (i)--(v), the following are equivalent:\\
(a) covariance under modular conjugations of modular automorphisms;\\
(b) covariance under modular conjugations of modular involutions;\\
(c) covariance under modular conjugations of wedge algebras; \\
(d) the modular conjugations $J(W)$ are representatives of $j(W)$ under a
representation of the covering group $\tPp$ of the proper Poincar\'e group,
under which the modular automorphisms are covariant;\\
(e) the modular conjugations $J(W)$ are the representatives of $j(W)$ under a
representation of the proper Poincar\'e group $\Pp$ satisfying the spectrum
condition, under which the vacuum is invariant, the modular conjugations,
modular automorphisms, modular involutions, and wedge algebras are all
covariant, and wedge duality holds;\\
(f) covariance under modular automorphisms of modular automorphisms;\\
(g) covariance under modular automorphisms of modular involutions;\\
(h) covariance under modular automorphisms of wedge algebras;\\
(i) the modular automorphisms $\Delta(W)^{it}$ are the representatives of
$\l(W,t)$ under a representation of the covering group $\tPr$ of the restricted
Poincar\'e group; \\
(j) the modular automorphisms $\Delta(W)^{it}$ are representatives of
$\l(W,t)$ under a unitary representation of the restricted Poincar\'e group
$\Pr$ satisfying the spectrum condition, under which the vacuum is invariant,
the modular conjugations, modular automorphisms, modular involutions, and
wedge algebras are all covariant, and wedge duality holds;\\
(k) the modular conjugations $J(W)$ are the representations of $j(W)$, and
the modular automorphisms $\Delta(W)^{it}$ of $\l(W,t)$, under a representation
of the proper Poincar\'e group $\Pp$ which represents orthochronous
transformations by unitary operators and time-reversing transformations by
antiunitary operators; this representation satisfies the spectrum condition,
under it the vacuum is invariant, the modular conjugations, modular
automorphisms, modular involutions, and wedge algebras are all covariant, and
wedge duality holds.\\
Assumption (v) is not necessary if any of (d)--(k) holds.
}

\smallskip\goodbreak
\noindent{\it Remark:}~~
{
The paper \cite{Ku1} assumes the existence and uniqueness of a covariant
representation of $\tPr$, and then shows essentially that (c) above implies
(e).  The paper \cite{GL2} shows essentially that (h) above implies (k).
}

\smallskip\goodbreak
\noindent{\it Proof:}~~
{
We have seen in the proof of Theorems 2 and 3 that wedge duality follows from
most of these statements.  Also, assumption (v) follows from (d), (e), (j),
or (k).  In fact, (k) implies all the other statements, (e) implies (a)--(d),
and (j) implies (f)--(i).  We will therefore begin by showing that (a)--(d) are
all equivalent to (e), continue by showing that (f)--(i) are all equivalent,
and finally show that (e), (i), (j), and (k) are all equivalent.

By Theorem 2, (c) is equivalent to (b), or to the conjunction of (a) above with
(a) of Theorem 4.  We show that (a) above implies (a) of Theorem 4.  If (a)
above holds, then by Lemma 5 we have $J(W_1)J(W_2)J(W_1)=V(W_1)J(j(W_1)W_2)$
where $V(W_1)$ depends only on $W_1$ and not on $W_2$.  But we may choose
$W_1=W_2$, from which we get $J(W_1)=V(W_1)J(j(W_1)W_1)=V(W_1)J(W_1)$ and
$V(W_1)=I$, so that modular conjugations are covariant under modular
conjugations.  Thus (a), (b), and (c) are all equivalent.  Clearly (d) implies
(a); on the other hand,
by Theorem 4, (a) of Theorem 4 combined with assumption
(v) implies that the modular conjugations generate a representation, which by
(c) is covariant.  Thus (d) is equivalent to (a)--(c).  That the spectrum
condition holds follows from a converse to Borchers' Theorem
\cite{Wb1,BGL2,Da}.   The equivalence of (a)--(e) then follows using Theorems 2
and 4.

By Theorem 3, (h) is equivalent to (g), or to the conjunction of (f) with
the covariance under modular automorphisms of modular conjugations.  We will
therefore first show that (f) implies the latter covariance.  If (f) holds,
then by Lemma 5 we have
\begin{equation}
\Delta(W_1)^{it}J(W_2)\Delta(W_1)^{-it}=V(W_1,t)J(\l(W_1,t)W_2)
\end{equation}
where $V(W_1,t)$ is independent of $W_2$.  We may choose $W_1=W_2$, obtaining
\begin{equation}
J(W_1)=\Delta(W_1)^{it}J(W_1)\Delta(W_1)^{-it}
=V(W_1,t)J(\l(W_1,t)W_1)=V(W_1,t)J(W_1),
\end{equation}
so that $V(W_1,t)=I$ identically.  From this follows the desired covariance
and the equivalence of (f), (g), and (h).  Clearly (i) implies (f), so we must
show that (f) implies (i).  This result is contained in \cite{BGL2}, cf. also
\cite{GL2}, and is analogous to that of Theorem 4.  We have immediately
one-parameter unitary groups $\Delta(W)^{it}$, which by the same procedure as
in Theorem 4 give a central weak Lie extension, and thus a unitary
representation of the covering group $\tPr$ under which the modular
automorphisms are covariant.  Since the vacuum is invariant under each
$\Delta(W)^{it}$, it is invariant under this representation.

Now we must show that (e), (i), (j), and (k) are all equivalent.  Let us begin
by showing that (e)
implies (i), (j) and (k).  In the representation of the
Poincar\'e group generated by the $J(W)$, let $\Delta_0(W)^{it}$ be the
representative of $\l(W,t)$.  Then the $\Delta(W)$ are covariant under the
$\Delta_0(W)^{it}$.  Thus for any particular wedge $W$, $\Delta(W)$ commutes
strongly with $\Delta_0(W)$, so the two positive operators have a common dense
set $D_\omega(W)$ of vectors $\psi$ such that $\Delta(W)^{iz}\psi$ and
$\Delta_0(W)^{iz}\psi$ are both entire analytic.  Let us take $W_1$, $W_2$ such
that $W_1\cap W_2\neq\emptyset$ and $W_1'\cap W_2'\neq\emptyset$.  For any
$\psi\in D_\omega(W_1)$, $\phi\in D_\omega(W_2)$ we may define a jointly entire
analytic function
\begin{equation}
f(z,w)=\ip<\Delta(W_1)^{i\bw}\Delta_0(W_1)^{i\bz}\psi|
\Delta(W_2)^{iw}\Delta_0(W_2)^{iz}\phi>\\
\end{equation}
satisfying
\begin{eqnarray}
|f(z,w)|&\leq&\left\|\Delta(W_1)^{i\bw}\Delta_0(W_1)^{i\bz}\psi\right\|\>
\left\|\Delta(W_2)^{iw}\Delta_0(W_2)^{iz}\phi\right\|\\
&=&\left\|\Delta(W_1)^{\Im w}\Delta_0(W_1)^{\Im z}\psi\right\|\>
\left\|\Delta(W_2)^{-\Im w}\Delta_0(W_2)^{-\Im z}\phi\right\|.
\end{eqnarray}
Then we may use Lemmas 5 and 6 to compute
\begin{eqnarray}
f(z,w+i/2)&=&
\ip<\Delta(W_1)^{1/2}\Delta(W_1)^{i\bw}\Delta_0(W_1)^{i\bz}\psi|
\Delta(W_2)^{-1/2}\Delta(W_2)^{iw}\Delta_0(W_2)^{iz}\phi>\nonumber\\
&=&\ip<J(W_1)\Delta(W_1)^{i\bw}\Delta_0(W_1)^{i\bz}\psi|
J(W_2)\Delta(W_2)^{iw}\Delta_0(W_2)^{iz}\phi>
\end{eqnarray}
and
\begin{eqnarray}
f(z+i/2,w)&=&
\ip<\Delta_0(W_1)^{1/2}\Delta(W_1)^{i\bw}\Delta_0(W_1)^{i\bz}\psi|
\Delta_0(W_2)^{-1/2}\Delta(W_2)^{iw}\Delta_0(W_2)^{iz}\phi>\nonumber\\
&=&\ip<J(W_1)\Delta(W_1)^{i\bw}\Delta_0(W_1)^{i\bz}\psi|
J(W_2)\Delta(W_2)^{iw}\Delta_0(W_2)^{iz}\phi>
\end{eqnarray}
from which we deduce that $f(z,w)=f(z+i/2,w-i/2)$.  Let us consider
$f(z+\z,w-\z)$ as a function of $\z$:  it is periodic in $\z$ with period
$i/2$, and satisfies a bound independent of $\Re\z$, so it is bounded
and, hence, constant.  Thus $f(z,w)=f(z+\z,w-\z)$ for all $z,w,\z$, and in
particular $f(t,0)=f(0,t)$, so that
\begin{equation}
\ip<\Delta(W_1)^{it}\psi|\Delta(W_2)^{it}\phi>
=\ip<\Delta_0(W_1)^{it}\psi|\Delta_0(W_2)^{it}\phi>.
\end{equation}
Since $\psi$ and $\phi$ may vary over dense sets, we conclude that for all
real $t$ we have
$\Delta(W_1)^{-it}\Delta(W_2)^{it}=\Delta_0(W_1)^{-it}\Delta_0(W_2)^{it}$, and
by suitably varying $W_1$ and $W_2$ we see that
$V(W,t)=\Delta_0(W)^{it}\Delta(W)^{-it}$ is in fact independent of $W$.  But
$V(W,t)$ is a one-parameter unitary group, so from $V(W,-t)=V(W',t)=V(W,t)$ we
find that $V(W,t)=1$ identically and $\Delta(W)=\Delta_0(W)$.  This implies
(i), (j), and (k) directly.

It will then suffice to show that (i) implies (a).  This result is contained
in \cite{BGL2}, and is closely connected with our Lemmas 5 and 6.  This
completes the proof.
}

\section{PCT and Spin-Statistics Theorems}

If $d$ is even, then the complete spacetime inversion is an element of
$\Pp$.  If the wedge algebras are covariant under a representation of $\Pp$
(or of $\tPp$) then the (antiunitary) representative of this inversion is
just the PCT operator $\Theta$.  Thus in even dimensions, Theorem 7 is a PCT
theorem; in odd dimensions, on the other hand, it seems that it must suffice to
have a representation of $\Pp$ (or of $\tPp$).

So far the argument has been stated entirely in terms of observable algebras,
and thus necessarily in terms of bosonic quantities.  In this case Theorem 7
guarantees representations of $\Pp$ and $\Pr$ rather than of their covering
groups---that is, it guarantees that all spins are integral.  Thus it is also
a spin-statistics theorems for bosonic statistics.  It may also be extended to
fermionic statistics by the use of a standard notation \cite{BW2}.  We let
$\Gamma$ be a unitary involution such that $\Gamma\Omega=\Omega$ and
$\Gamma\A(W)\Gamma=\A(W)$ for all $W$.  Operators that commute with
$\Gamma$ are intended to be bosonic, while those that anticommute are to
be fermionic.  Let $Z=(I+i\Gamma)/(1+i)$, and let us alter assumption (i) as
follows:\\
(i$'$) $Z\A(W')Z^*\subset\A(W)'$ (twisted locality).\\
Wedge duality is likewise altered to $Z\A(W')Z^*=\A(W)'$ (twisted wedge
duality).  The alterations correspond to normal commutation relations:
commutation between spacelike separated operators, except that two spacelike
separated fermionic operators anticommute.  The argument proceeds much as
before, save that in place of covariance under the modular conjugations we must
substitute covariance under the twisted modular conjugation operators
$Z^*J(W)$.  $\Gamma$ commutes with every $J(W)$ and $\Delta(W)$, so
$Z^*J(W)=J(W)Z$ is again an antiunitary involution.  The results are as before,
except that (i$'$) implies that $Z^*J(W_R)Z=J(W_L)$, so that the final argument
in Theorem 4 now shows that $R(2\pi)=Z^*J(W_R)ZJ(W_L)=Z^2=\Gamma$.  Thus in
each case we have representations of the covering groups $\tPp$ or $\tPr$, but
subject to the condition that $R(2\pi)=\Gamma$.  The modified Theorem 7$'$ as
follows is therefore an algebraic PCT and spin-statistics theorem.

\bigskip\goodbreak
\noindent{\bf Theorem 7$'$:}~~
{\em
Under assumptions (i$'$) and (ii)--(iv), any of the subparts corresponding to
(a)--(j) of Theorem 7 is equivalent to the following:\\
(k$'$) the twisted modular conjugations $Z^*J(W)$ are the representations of
$j(W)$, and the modular automorphisms $\Delta(W)^{it}$ of $\l(W,t)$, under a
representation of the covering group $\tPp$ of the proper Poincar\'e group,
subject to the condition that $R(2\pi)=\Gamma$, which represents orthochronous
transformations by unitary operators and time-reversing transformations by
antiunitary operators; this representation satisfies the spectrum condition,
and under it the vacuum is invariant, and the modular conjugations, modular
automorphisms, modular involutions, and wedge algebras are all covariant, and
twisted wedge duality holds.\\
}
\bigskip\goodbreak

However, there is more than this that can be said.  Using Theorem 7$'$
we can show only that if any one of the equivalent conditions (a$'$)--(k$'$)
holds for the field net, then so also do all the rest.  What is in fact true,
with only a few additional assumptions, is that modular covariance for the
observable net implies modular covariance for the field net.  However, the
present context does not appear to be the appropriate one for a discussion
of these issues.  Various aspects of the matter are treated in \cite{GL1} and
in \cite{BGL1,GL2,Ku1,GL3}.  In some sense this is the true spin-statistics
theorem in this context, but of course it depends on the modular covariance of
the observable net, precisely the question studied here so far.  It is for this
reason, as well as for clarity of exposition, that the presentation in the
previous sections has been entirely in terms of observables:  modular
covariance for the observables implies modular covariance generally.  The
remainder of this paper will discuss modular covariance for field nets
directly.

\section{Localized States for Elementary Systems}

We must now explain our earlier statement that these theorems allow
us to reduce the study of the possible modular structures of nets to the study
of nets without interaction; this will give substance to the rather abstract
results of Theorems 7 and 7$'$.  We first note that the relevant portions of
these theorems apply equally to families of `modular operators' not necessarily
associated with any von Neumann algebras.  Let us consider a family of
antiunitary involutions $J(W)$ and unbounded positive operators $\Delta(W)$, or
equivalently the corresponding unbounded antilinear involutions
$S(W)=J(W)\Delta(W)^{1/2}$, acting on a Hilbert space $\H_1$.   There is then a
corresponding family of subsets $R(W)$ of $\H_1$ defined by
$R(W)=\[\psi|S(W)\psi=\psi]$.  Conversely, the operators can be recovered from
the $R(W)$ by letting $S(W)(\psi+i\phi)=\psi-i\phi$ for every
$\psi,\phi\in R(W)$.  Consider the following assumptions:\\
(i) $S(W)^*\psi=\psi$ for every $\psi\in R(W')$;\\
(ii) $R(W_1)\subset R(W_2)$ whenever $W_1\subset W_2$;\\
(iii) $R_{12}=R(W_1)\cap R(W_2)$ is such that $R_{12}+iR_{12}$ is dense in
$\H_1$ whenever $W_1\cap W_2\neq\emptyset$;
(iv) for all $W_1$ and $W_2$, $R(W_1)=\[R(W_1)\cap R(W)|{W\|_sW_2}]^-$;\\
(v) $J(\Lambda W)$ is a weakly continuous function of $\Lambda\in\Pr$;\\
or, in the more general case, given a unitary involution $\Gamma$, \\
(i$'$) $ZS(W)^*\psi=\psi$ for every $\psi\in R(W')$.\\
These correspond directly to assumptions (i)--(v) and (i$'$) for
families of wedge algebras $\A(W)$.  Wedge duality corresponds to
$S(W')=S(W)^*$, and twisted wedge duality to $S(W')=ZS(W)^*$.  Theorems 7 and
7$'$ still hold, with the omission of (c), (h), and all other references to
wedge algebras.  Thus it is still reasonable to speak of modular covariance for
such a family of modular operators.  Notice that the $\Delta(W)$ do not
determine the $J(W)$ uniquely, but only up to a phase operator $V$.  This is
entirely consistent with the results of Lemma 5.

Next, we point out that a construction analogous to that of the free fields
can produce a family of wedge algebras describing non-interacting particles
corresponding to any such family of modular operators.  Let us write
$\H_1=\H_1^b\oplus\H_1^f$ where $\H_1^b$ and $\H_1^f$ are the eigenspaces of
$\Gamma$ with eigenvalues $+1$ and $-1$ respectively (the bosonic and
fermionic one-particle spaces).  Then over $\H_1$ we construct a mixed
bosonic/fermionic Fock space $\H=\H^b\otimes\H^f$, where $\H^b$ is the
bosonic (symmetric) Fock space over $\H_1^b$, and $\H^f$ is the fermionic
(antisymmetric) Fock space over $\H_1^f$.  On $\H$ it is possible to define
bosonic field operators $\phi^b(\psi)$ for $\psi\in\H_1^b$, fermionic field
operators $\phi^f(\psi)$ for $\psi\in\H_1^f$, and, by adding these, general
self-adjoint field operators $\phi(\psi)$ for $\psi\in\H_1$, defined such that
$\phi(\psi)\Omega=\psi$.  We then define algebras $\A(W)$ generated by the
field operators $\phi(\psi)$ for every $\psi\in R(W)$.  If the $R(W)$ satisfy
conditions (i)--(v) or (i$'$) above, then the $\A(W)$ satisfy the corresponding
conditions (i)--(v) or (i$'$) previously defined for them.  Furthermore the
modular conjugations and automorphisms for the $\A(W)$ agree with the specified
$J(W)$ and $\Delta(W)$ on the one-particle space in $\H$, which we may identify
with $\H_1$.  Such a family of wedge algebras describes a system without
interaction, but it clearly gives examples of any phenomenon that occurs at the
level of modular operators.  If the $J(W)$ and $\Delta(W)$ satisfy the
conditions of Theorem 7$'$, then the $\A(W)$ correspond to a (generalized) free
field, and thus certainly arise from a net of local algebras.  If they do not,
however, then by the Bisognano-Wichmann theorem they cannot correspond to any
set of Wightman fields, and they may or may not arise from a local net.  We
knew already that every family of wedge algebras produces a corresponding
family of modular operators; what this shows is that a family of modular
operators produces a family of wedge algebras.  Thus for every result about
families of wedge algebras, there is a corresponding result about families of
modular operators, and vice versa.  The study of modular structures for general
families of wedge algebras is reduced to that of modular structures in the
abstract, which correspond in this way to algebras without interaction.

Systems of algebras of this type are not the only ones to describe models
without interaction, but they form an important class:  given any one-particle
space $\H_1$, these are those that describe all states of arbitrarily many such
particles, present together without interaction.  The Hilbert space $\H$ is
uniquely determined by $\H_1$ and $\Gamma$, but the algebras $\A(W)$ depend on
the family of modular operators on $\H_1$---this family describes the
localization properties of the one-particle states, and we will refer to it as
the localization structure for such a model.  For the free fields, the
localization structure is determined entirely by the representation of $\tPp$,
by the Bisognano-Wichmann condition, but it is not known whether there might be
other possibile localization structures not corresponding to free fields.  Here
we study the question of uniqueness for families of modular operators on
$\H_1$, assuming the existence of an appropriate representation of the
Poincar\'e group.  In this case there is a distinguished family
$J_0(W)=U(j(W))$, $\Delta_0(W)^{it}=U(\l(W,t))$ of modular operators.  For any
other family $J(W)$, $\Delta(W)$ covariant under the representation, we may
define $J'(W)=J(W)J_0(W)=J_0(W)J(W)$ and
$\Delta'(W)^{it}=\Delta(W)^{it}\Delta_0(W)^{-it}$, taking advantage of the
commutation properties provided by the covariance.  If our representation is
only of the restricted Poincar\'e group, then we may at least define
$\Delta_0(W)$ and $\Delta'(W)$ in the same way.

The most interesting case is that in which $\H_1$ carries an irreducible
representation of the Poincar\'e group, corresponding to Wigner's notion
of an elementary system.  Newton and Wigner \cite{NW} studied the possibilities
for localization of states in the traditional quantum-mechanical sense on
elementary systems; what we are studying here is a different sort of
localization structure, one appropriate to the systems we describe here, and
in particular to the free fields.  We will treat not only elementary systems,
but also reducible representations, provided they satisfy certain multiplicity
restrictions.

\bigskip\goodbreak
\noindent{\bf Lemma 8:}~~
{\em
Suppose assumptions (i$'$), (ii), and (iii) hold.  Let $U(\lambda)$ be a
representation of the covering group $\tPr$ of the restricted Poincar\'e group,
satisfying the spectrum condition, under which the modular automorphisms
$\Delta(W)$ are covariant.  For any two wedges $W_1, W_2$ such that
$W_1\cap W_2\neq\emptyset$ and $W_1'\cap W_2'\neq\emptyset$, there is a unitary
operator $V(W_1,W_2)$ such that
\begin{equation}
\ip<\Delta'(W_1)^{1/2}\psi|\Delta'(W_2)^{-1/2}\phi>=\<\psi|V(W_1,W_2)\phi>
\end{equation}
for every $\psi\in D(\Delta'(W_1)^{1/2})$, $\phi\in D(\Delta'(W_2)^{-1/2})$
(and each of these sets is dense).
}

\smallskip\goodbreak
\noindent{\it Proof:}~~
{
Since the $\Delta(W)$ are covariant under the $\Delta_0(W)^{it}$, we see that
for any particular wedge $W$, $\Delta_0(W)$ commutes strongly with $\Delta(W)$.
Thus there is a dense domain $D_\omega(W)$ on which
$\Delta'(W)^{iz}=\Delta(W)^{iz}\Delta_0(W)^{-iz}$ for all complex $z$.
Thus for all $\psi\in D_\omega(W_1)$, $\phi\in D_\omega(W_2)$, we may use
Lemmas 5 and 6 (appropriately extended for the possibility of fermions) to
compute
\begin{eqnarray}
\ip<\Delta'(W_1)^{1/2}\psi|\Delta'(W_2)^{-1/2}\phi>&=&
\ip<\Delta_0(W_1)^{-1/2}\Delta(W_1)^{1/2}\psi|
   \Delta_0(W_2)^{1/2}\Delta(W_2)^{-1/2}\phi>\nonumber\\
&=&\ip<\Delta(W_1)^{1/2}\psi|U(j(W_1)j(W_2))\Delta(W_2)^{-1/2}\phi>\\
&=&\ip<\Delta(W_1)^{1/2}\psi|
   \Delta(j(W_1)W_2')^{-1/2}U(j(W_1)j(W_2))\phi>\nonumber\\
&=&\ip<\psi|J(W_1)U(j(W_1)j(W_2))J(W_2)\phi>,\nonumber
\end{eqnarray}
and the operator in the last expression is unitary.  The result then follows by
linearity for all $\psi$, $\phi$ for which the expression is defined.
}
\bigskip\goodbreak

Next, we see that relations of this sort cannot be satisfied by bounded
operators, by commuting operators, or even by matrices of commuting
operators.

\bigskip\goodbreak
\noindent{\bf Lemma 9:}~~
{\em
If the equality of Lemma 8 holds, and if each $\Delta'(W)$ is a bounded
operator, then $\Delta'(W)=I$ for every wedge $W$.
}

\smallskip\goodbreak
\noindent{\it Proof:}~~
{
In this case, $\Delta'(W_1)^{1/2}\Delta'(W_2)^{-1/2}=V(W_1,W_2)$ whenever
$W_1$, $W_2$ are as in Lemma 8.  Thus
$V(W_1,W_2)^*=\Delta'(W_2)^{-1/2}\Delta'(W_1)^{1/2}$, so that
$V(W_1,W_2)V(W_1,W_2)^*=\Delta'(W_1)^{1/2}\Delta'(W_2)^{-1}\Delta'(W_1)^{1/2}
=I$.  Thus $\Delta'(W_2)^{-1}=\Delta'(W_1)^{-1}$, and using appropriate pairs
of wedges we see that $\Delta'(W)$ is independent of $W$.  But then
$\Delta'(W)^{it}=\Delta'(W')^{-it}=\Delta'(W)^{-it}$ so that $\Delta'(W)=I$.
}

\bigskip\goodbreak
\noindent{\bf Lemma 10:}~~
{\em
If the equality of Lemma 8 holds, and if there is an abelian von Neumann
algebra $\N$ and an embedding of the $n\times n$ matrix algebra $M_n(\N)$ over
$\N$ in $\B(\H)$ such that every $\Delta'(W)^{it}$ lies in $M_n(\N)$, then
$\Delta'(W)=I$ for every wedge $W$.
}

\smallskip\goodbreak
\noindent{\it Proof:}~~
{
For simplicity we may consider $\N$ as generated by a single self-adjoint
operator $X$.  For any unbounded measurable function $f(X)$, the sets
$E_a=f^{-1}([-a,a])$ form an increasing family of measurable sets on which $f$
is bounded, and such that $\cup_a E_a=\RR$.  Likewise if we have a finite
family $f_i$ of such functions, then $E_a=\cap_i f_i^{-1}([-a,a])$ has the
same properties.  For each wedge $W$, $\Delta'(W)$ is an $n\times n$ matrix
whose entries are unbounded measurable functions of $X$.  Since there are only
finitely many
such entries, for any finite collection of wedges $W$ there is
such a family of sets $E_a$ on which every entry in every $\Delta'(W)$ is
finite.  If $\Pi_a$ are the corresponding spectral projections for $X$, then
$\Delta'_a(W)=\Pi_a\Delta'(W)=\Delta'(W)\Pi_a$ is in $M_n(\N)$ for every $a$,
and $\Pi_a$
tends strongly to the identity as $a\rightarrow\infty$.  Then
if $W_1$ and $W_2$ are as in Lemma 8, $\Delta'_a(W_1)$ and $\Delta'_a(W_2)$
satisfy the relation of Lemma 8, but are both bounded, and hence the
proof of Lemma 9 (which requires only a suitable finite collection of wedges)
can be used to show that $\Delta'_a(W)$ is the identity on $\Pi_a\H$.  Thus in
the limit we have $\Delta'(W)=I$.
}
\bigskip\goodbreak

These results imply uniqueness in case the representation of the Poincar\'e
group satisfies certain multiplicity conditions.  These will not hold on the
full Fock space, but they can be satisfied on the one-particle space.  In
particular, they hold in the case of elementary systems.  These conditions also
suffice to guarantee the uniqueness of the representation of the Poincar\'e
group under which the net is covariant.

\bigskip\goodbreak
\noindent{\bf Theorem 11:}~~
{\em
If $\H_1$ carries a representation $U(\l)$ of the covering group $\tPr$ of the
restricted Poincar\'e group, satisfying the spectrum condition, and such that
for any given mass there occur only finitely many finite-spin irreducible
representations, each with finite multiplicity, then there is up to unitary
equivalence at most one set of modular operators covariant under the $U(\l)$
and satisfying assumptions (i$'$), (ii), and (iii).  It must satisfy
assumptions (iv) and (v), twisted wedge duality, and modular covariance.  The
modular automorphisms are uniquely determined, while the modular conjugations
are determined only up to a unitary operator commuting with all $U(\l)$.
}

\smallskip\goodbreak
\noindent{\it Proof:}~~
{
There is a set of free fields with $\H_1$ for one-particle space if and
only if $\H_1$ admits a PCT operator with respect to the given representation
$U(\l)$, which thus extends to a representation of the proper group $\tPp$.
Let us first assume that this is the case, so that there is a
free-field net satisfying (i$'$) and (ii)--(v), twisted wedge duality, and
modular covariance, for which $\H_1$ is the one-particle space.  The
PCT operator and the representation of $\tPp$ are unique up to a
Poincar\'e-invariant unitary operator.  By the Bisognano-Wichmann results, the
modular operators on $\H_1$ for any free-field net must come from one of these
representations.

Let us assume that there is some other set $J(W)$, $\Delta(W)$ of modular
operators on $\H_1$, and let us use the notation of Lemma 8.  We now require a
version of Borchers' Theorem \cite{Bo1}; one directly applicable to the present
situation may be found in Theorem 3 of \cite{Da}.  This result, dependent on
the spectrum condition, implies that
$\Delta(W)^{it}T(x)\Delta(W)^{-it}=\Delta_0(W)^{it}T(x)\Delta_0(W)^{-it}$
if $T(x)$ is any translation.  Thus $\Delta'(W)$ commutes strongly with
all translations, and with the von Neumann algebra $\T$ generated by all
translations.  Since $\Delta_0(W)$ and $\Delta'(W)$ commute with the mass
operator, so also does $\Delta(W)$, and without loss of generality we may
restrict ourselves to the case of a single mass $m$.  Then for this case we
note that the multiplicity conditions imply that $\T'$ is isomorphic to
$M_n(\T)$, where $n$ is the total number of local degrees of freedom for all
particles of mass $m$.  Then Lemma 10 gives us our result immediately.

On the other hand, if $\H_1$ does not admit a PCT operator, then we may
substitute for it a direct sum $\H_1\oplus\H_1'$, where $\H_1'$ is
the PCT conjugate of $\H_1$, without disturbing the multiplicity conditions,
to obtain a representation which does admit a PCT operator.  Any set of modular
operators on $\H_1$ gives a set of modular operators on $\H_1\oplus\H_1'$ by
the same operation.  But by the argument just given, there is precisely one set
of modular operators for $\H_1\oplus\H_1'$, and it satisfies modular
covariance.  Thus there can be no set of modular operators for $\H_1$.
}
\bigskip\goodbreak

Thus if the particle content of $\H_1$ is sufficiently restricted, there is
only one possible localization structure, that corresponding to the usual free
fields.  If we consider non-interacting systems of the sort described above,
the only possibilities are those corresponding to free fields, which
necessarily arise from local nets.  Thus in these cases the Bisognano-Wichmann
condition, wedge duality, the PCT theorem, and the spin-statistics theorem must
hold.  As the examples of \cite{St} and \cite{La} show, the net of modular
operators and the representation of the Poincar\'e group need not be unique in
the absence of multiplicity constraints.  These examples still satisfy modular
covariance, but it is not known whether there might be other
Poincar\'e-covariant structures not satisfying modular covariance.  If we
omit the hypothesis of Poincar\'e covariance, there will be many possibilities
not satisfying wedge duality \cite{Yn}.  One additional interesting possibility
suggested by the result of \cite{Ku2} is that the modular automorphism group
might have a geometric interpretation differing from that of Bisognano and
Wichmann by a translation parallel to the vertex of the wedge.  What we see
here is that this possibility cannot occur in simple non-interacting models.

\section{Asymptotic Locality}

In this section we will assume that we are dealing with a Poincar\'e-covariant
net of local algebras having a complete asymptotic interpretation via the
Haag-Ruelle scattering theory, in terms of massive particles.  Standard
assumptions for nets of local algebras then imply that the wedge algebras must
satisfy conditions (i$'$), (ii), and (iii).  The restriction to massive
particles is probably not necessary, but the scattering theory for massive
particles is considerably simpler.  The Haag-Ruelle theory assures us that the
non-interacting behavior of the asymptotic particles is described by
non-interacting systems of the sort constructed in the last section, but it
does not specify any particular localization structure.  The localization
structure for the non-interacting systems must be determined from the
interacting net after the manner described in \cite{La} (asymptotic locality).
However, if the theory is such that the asymptotic one-particle space $\H_1$
satisfies the multiplicity conditions of Theorem 11, then we know that the only
possible localization structures describing the free behavior of the asymptotic
particles are those of free-field nets.  There are in fact two sets of free
fields relevant to scattering, the in-fields and the out-fields, and these
differ by a Poincar\'e-invariant unitary operator, the S-matrix.

\bigskip\goodbreak
\noindent{\bf Theorem 12:}~~
{\em
Let $\A(W)$ be a family of wedge algebras derived from a local net covariant
under a
representation $U(\l)$ of the covering group $\tPr$ of the restricted
Poincar\'e group, that satisfies the spectrum condition.  Suppose the $\A(W)$
satisfy conditions (i$'$), (ii), and (iii), and also asymptotic completeness,
with a one-particle subspace $\H_1$ on which the mass spectrum is discrete and
positive, and for which the multiplicity conditions of Theorem 11 hold.  Then
the $\A(W)$ satisfy twisted wedge duality and modular covariance.
}

\smallskip\goodbreak
\noindent{\it Proof:}~~
{
We will use the same notation as in Theorem 11.  As in Theorem 11, all the
modular operators for wedge regions commute with the mass.  The one-particle
space $\H_1$ can be distinguished as the discrete-mass subspace, so that
each $\Delta(W)^{it}$ must leave $\H_1$ invariant, and must restrict to a set
of modular operators on $\H_1$ satisfying conditions (i$'$), (ii), and (iii).
By Theorem 11, we see that $\H_1$ must admit a PCT operator, and $\Delta(W)$
must agree with $\Delta_0(W)$ on $\H_1$.  The modular conjugations may differ
on $\H_1$, but only by a Poincar\'e-invariant unitary operator.  This is
consistent with the asymptotic locality result of \cite{La}.  What we now wish
to show is that the same result holds on all of $\H$.  We will do this by
showing that the modular operators act multiplicatively on the asymptotic
fields.  Then the modular automorphisms must arise from the common
representation of the Poincar\'e group, and both twisted wedge duality and
modular covariance must hold.

We first notice that by the Tomita-Takesaki theorem, for any wedge $W_0$, we
have
\begin{equation}
\Delta(W_0)^{it}\A(W_0)\Delta(W_0)^{-it}=\A(W_0)
=\Delta_0(W_0)^{it}\A(W_0)\Delta_0(W_0)^{-it}
\end{equation}
for all real $t$.  Also, by Borchers' Theorem \cite{Bo1}, for every translation
$T(x)$ we have
\begin{equation}
\Delta(W_0)^{it}T(x)\Delta(W_0)^{-it}=\Delta_0(W_0)T(x)\Delta_0(W_0)^{-it}
\end{equation}
for all $t$.  Thus by covariance we have
\begin{equation}
\Delta(W_0)^{it}\A(W_1)\Delta(W_0)^{-it}
=\Delta_0(W_0)^{it}\A(W_1)\Delta_0(W_0)^{-it}
\end{equation}
for all real $t$ and any $W_1\|_s W_0$, and by the same reasoning likewise
\begin{equation}
J(W_0)\A(W_1)J(W_0)=J_0(W_0)\A(W_1)J_0(W_0)
\end{equation}
for every $W_1\|_s W_0$.  The collection of $\A(W)$ and $Z^*\A(W)'Z$ for each
$W\|_s W_0$
forms a two-dimensional net of wedge algebras, and the set of
algebras $\A(W_1)\cap Z^*\A(W_2)'Z$ for $W_1,W_2\|_s W_0$ forms a
two-dimensional local net.  It is highly degenerate, but it is still possible
to construct its Haag-Ruelle scattering theory.  The asymptotic fields will be
generalized free fields, produced by the same sort of dimension-reducing
procedure from the original asymptotic fields.  Their one-particle space will
again be $\H_1$, but now reinterpreted as carrying a highly reducible
representation of the two-dimensional Poincar\'e group.  With respect to this
net, $\Delta'(W_0)^{it}$ and $J'(W_0)$ are local internal symmetries, and as in
\cite{La}, they must act multiplicatively on the asymptotic fields.  But we
have already concluded in the previous paragraph that $\Delta'(W_0)$ is trivial
on $\H_1$, and $J'(W_0)$ is a Poincar\'e-invariant unitary operator, so this
must also be true on all of $\H$.
}
\bigskip\goodbreak

Thus we see that the Bisognano-Wichmann condition, modular covariance, wedge
duality, the PCT theorem, and the spin-statistics theorem all hold for nets
satisfying asymptotic completeness with appropriate restrictions on their
asymptotic particle content:  namely, that the particle spectrum be discrete
and positive in mass, and finite in spin and total multiplicity for each mass.

\bigskip\bigskip\goodbreak
\noindent{\Large\bf Acknowledgements}
\bigskip

\noindent{
I would like to thank Sergio Doplicher and the Universit\`a di Roma for their
hospitality while this work was in progress, and Daniele Guido and Bernd
Kuckert for their invaluable help and inspiration.
}


\begin{thebibliography}{99}
\bibitem{BW1} Bisognano, J.~J, Wichmann, E.~H.:  On the Duality Condition
for a Hermitian Scalar Field.  J. Math. Phys. {\bf 16}, 985 (1975).
\bibitem{BW2} Bisognano, J.~J, Wichmann, E.~H.:  On the Duality Condition
for Quantum Fields.  J. Math. Phys. {\bf 17}, 303 (1976).
\bibitem{Bo1} Borchers, H.~J.:  The CPT-Theorem in Two-Dimensional Theories of
Local Observables.  Commun. Math. Phys. {\bf 143}, 315 (1992).
\bibitem{Bo2} Borchers, H.~J.:  On Modular Inclusion and Spectrum Condition.
Lett. Math. Phys. {\bf 27}, 311 (1993).
\bibitem{BGL1} Brunetti, R., Guido, D., Longo, R.:  Modular Structure and
Duality in Conformal
Quantum Field Theory.  Commun. Math. Phys. {\bf 156}, 201
(1993).
\bibitem{BGL2}  Brunetti, R., Guido, D., Longo, R.:  Group Cohomology, Modular
Theory, and Spacetime Symmetries.  Rev. Math. Phys. {\bf 7}, 57 (1995).
\bibitem{BS} Buchholz, D., Summers, S.~J.:  An Algebraic Characterization of
Vacuum States in Minkowski Space.  Commun. Math. Phys. {\bf 155}, 459 (1993).
\bibitem{Bg} Burgoyne, N.:  On the Connection of Spin with Statistics.
Nuovo Cimento {\bf 8}, 807 (1958).
\bibitem{Da} Davidson, D.~R.:  Endomorphism Semigroups and Lightlike
Translations.  To appear in Lett. Math. Phys.
\bibitem{FJ} Fredenhagen, K., J\"orss, M.:  Conformal Haag-Kastler Nets,
Poinlike Localized Fields and the Existence of Operator Product Expansions.
Preprint (1994).
\bibitem{FG} Fr\"ohlich, J., Gabbiani, F.:  Operator Algebras and Conformal
Field Theory.  Commun. Math. Phys. {\bf }155, 569 (1993).
\bibitem{GL1} Guido, D., Longo, R.:  Relativistic Invariance and Charge
Conjugation in Quantum Field Theory.  Commun. Math. Phys. {\bf 148}, 521
(1992).
\bibitem{GL2} Guido, D., Longo, R.:  An Algebraic Spin and Statistics
Theorem I.  Commun. Math. Phys. {\bf 172}, 517 (1995).
\bibitem{GL3} Guido, D., Longo, R.:  A Conformal Spin and Statistics
Theorem.  Preprint (1995).
\bibitem{Jo} Jost, R.:  Eine Bemerkung zum CTP Theorem.  Helv. Phys. Acta
{\bf 30}, 409 (1957).
\bibitem{Ku1} Kuckert, B.:  A New Approach to Spin and Statistics.  Preprint
(1995).
\bibitem{Ku2} Kuckert, B.:  Borchers' Commutation Relations and Modular
Symmetries.  Preprint (1995).
\bibitem{La} Landau, L.~J.:  Asymptotic Locality and the Structure of Local
Internal Symmetries.  Commun. Math. Phys. {\bf 17}, 156 (1970).
\bibitem{LZ} L\"uders, G., Zumino, B.:  Connection between Spin and
Statistics.  Phys. Rev. {\bf 110}, 1450 (1958).
\bibitem{NW} Newton, T.~D., Wigner, E.~P.:  Localized States for Elementary
Systems.  Rev. Mod. Phys. {\bf 21}, 400 (1949).
\bibitem{St} Streater, R.~F.:  Local Fields with Wrong Connection Between Spin
and Statistics.  Commun. Math. Phys. {\bf 5}, 88 (1967).
\bibitem{Wb1} Wiesbrock, H.~W.:  A Comment on a Recent Work of Borchers.
Lett. Math. Phys. {\bf 25}, 157 (1992).
\bibitem{Wb2} Wiesbrock, H.~W.:  Symmetries and Half-Sided Modular Inclusions
of von
Neumann Algebras.  Lett. Math. Phys. {\bf 28}, 107 (1993).
\bibitem{Wb3} Wiesbrock, H.~W.:  Half-Sided Modular Inclusions of von Neumann
Algebras.  Commun. Math. Phys. {\bf 157}, 83 (1993).
\bibitem{Wb4} Wiesbrock, H.~W.:  Conformal Quantum Field Theory and Half-Sided
Modular Inclusions of von Neumann Algebras.  Commun. Math. Phys. {\bf 158}, 537
(1993).
\bibitem{Yn} Yngvason, J.:  A Note on Essential Duality.  Preprint (1993).

\end{thebibliography}
\end{document}